\documentclass[twocolumn]{aastex63}

\usepackage{color}
\newcommand{\p}[1]{{\color{magenta}{#1}}}
\newcommand{\degree}{^\circ} 
\newcommand{\BE}{\begin{equation}}
\newcommand{\EE}{\end{equation}}
\newcommand{\BA}{\begin{eqnarray}}
\newcommand{\EA}{\end{eqnarray}}
\newcommand{\Fig}[1]{Figure~\ref{fig_#1}}
\renewcommand{\fig}[1]{Figure~\ref{fig_#1}}

\newcommand{\figs}[2]{Figures~\ref{fig_#1} and \ref{fig_#2}}

\newcommand{\sect}[1]{Section~\ref{sect_#1}}
\newcommand{\app}[1]{Appendix~\ref{app_#1}}

\newcommand{\eg}{{\it e.g.},}

\newcommand{\ie}{{\it i.e.}}

\newcommand{\phia}{\phi^{\rm M}_{\rm a}}

\newcommand{\phic}{\phi^{\rm M}_{\rm c}}
\newcommand{\phiwl}{\phi^{\rm WL}_{\rm U}}
\newcommand{\phiwlm}{\phi^{\rm WLM}_{\rm U}}

\received{***}
\revised{***}
\accepted{***} %{\today}
\submitjournal{ApJ}

\shorttitle{AR Tilt Angles from White-Light Images and Magnetograms}
\shortauthors{Poisson et al.}
\begin{document}

\title{Active-Region Tilt Angles from White-Light Images and Magnetograms: The Role of Magnetic Tongues}

\correspondingauthor{Mariano Poisson}
\email{mpoisson@iafe.uba.ar}

\author[0000-0002-4300-0954]{Mariano Poisson}
\affiliation{Instituto de Astronom\'\i a y F\'\i sica del Espacio, IAFE, CONICET-UBA, CC. 67, Suc. 28, 1428 Buenos Aires, Argentina}

\author[0000-0001-8215-6532]{Pascal D\'emoulin}
\email{Pascal.Demoulin@obspm.fr}
\affiliation{LESIA, Observatoire de Paris, Universit\'e PSL, CNRS, Sorbonne Universit\'e, Univ. Paris Diderot, Sorbonne Paris Cit\'e, 5 place Jules Janssen, 92195 Meudon, France}

\author[0000-0001-9311-678X]{Cristina H. Mandrini}
\email{mandrini@iafe.uba.ar}
\affiliation{Instituto de Astronom\'\i a y F\'\i sica del Espacio, IAFE, CONICET-UBA, CC. 67, Suc. 28, 1428 Buenos Aires, Argentina}
\affiliation{Universidad de Buenos Aires, Facultad de Ciencias Exactas y Naturales, 1428 Buenos Aires, Argentina}
%\nocollaboration{1}

\author[0000-0001-8830-4022]{Marcelo C. L\'opez Fuentes}
\email{lopezf@iafe.uba.ar}
\affiliation{Instituto de Astronom\'\i a y F\'\i sica del Espacio, IAFE, CONICET-UBA, CC. 67, Suc. 28, 1428 Buenos Aires, Argentina}


\begin{abstract}

% Astrophysical Journal (ApJ),has a 250 word limit for the abstract
% context
The presence of elongations in active region (AR) polarities, called magnetic tongues, are mostly visible during their emergence phase.
% and affect the measurement of several characteristics; in particular, AR tilt angles.  
% aims 
AR tilts have been measured thoroughly using long-term white-light (WL) databases, sometimes combined with magnetic field information. Since the influence of magnetic tongues on WL tilt measurements has not been taken into account before, we aim to investigate their role in tilt-angle values and to compare them with those derived from LOS magnetograms.
% Methods 
We apply four methods to compute the tilt angle of generally bipolar ARs: one applies the k-means algorithm to WL data, a second one includes the magnetic field sign of the polarities to WL data, and a third one uses the magnetic flux-weighted center of each polarity. The tilt values computed in any of these ways are affected by the presence of magnetic tongues. Therefore, we apply the newly developed Core Field Fit Estimator (CoFFE) method to separate the magnetic flux in the tongues from that in the AR core.
%  results
We compare the four computed tilt-angle values, as well as these with the ones reported in long-term WL databases. For ARs with low magnetic flux tongues the different methods report consistent tilt-angle values. But for ARs with high flux tongues there are noticeable discrepancies between all methods indicating that magnetic tongues affect differently WL and magnetic data. However, in general, CoFFE achieves a better estimation of the main bipole tilt because it removes both the effect of tongues as well as the emergence of secondary bipoles { when it occurs in between the main bipole magnetic polarities.} 
\end{abstract}

\keywords{Solar magnetic flux emergence --- Solar active regions --- Solar active region magnetic fields --- Bipolar sunspot groups}
% keywords from http://astrothesaurus.org/

%\keywords{Physical data and processes: magnetic fields, Sun: photosphere, Sun: magnetic fields}
   
%%%%%%%%%%%%%%%%%%%%%%%%%%%%%%%%%%%%%%%%%%%%%%%%%%%%%%%%%%%%%%%%%%%%%%%%%%%%%%%%%%%%%%
\section{Introduction} 
\label{sect_Introduction}
%%%%%%%%%%%%%%%%%%%%%%%%%%%%%%%%%%%%%%%%%%%%%%%%%%%%%%%%%%%%%%%%%%%%%%%%%%%%%%%%%%%%%%

%{\S\bf --- ARs as emerging flux ropes} \\
 The simplest manifestation of an active region (AR) is in the form of a magnetic bipolar configuration, \ie\ made of a main positive and a main negative polarity \citep{vanDriel15}. 
{ Furthermore, ARs with two sunspots or sunspot groups of opposite magnetic polarity, called a $\beta$ configuration, form the large majority of ARs all along the solar cycle \citep{Jaeggli16}; therefore, it is worth to study and understand general AR properties using mainly bipolar ARs.} 
Large sets of observational data, theoretical developments \citep[\ie~dynamo models, see \eg\  the reviews of][and references therein]{Charbonneau14,Brun17}, as well as magnetohydrodynamic (MHD) simulations \citep[see the reviews by][and references therein]{Fan09,Cheung14,Toriumi14}, support the idea that bipolar active regions are the consequence of the emergence of magnetic flux tubes. These flux tubes, which have been called $\Omega$ loops \citep{Zwaan87}, { originate in the toroidal magnetic field created by the dynamo mechanism in the convection zone}. Their field is amplified and deformed by differential rotation and convective motions until they become buoyant and emerge in the form of twisted flux-tubes or flux ropes { \citep[FRs,][]{Fan09r,Nelson13}}. However, other MHD simulations explain the formation of ARs due to the local amplification and structuring of the magnetic field in the upper layers of the convective zone \citep[see the review by][]{Brandenburg18}.  

%{\S\bf --- Tilt as a consequence of Coriolis force action} \\
As the toroidal magnetic flux rises through the convection zone, the Coriolis force acts on the FRs so that they emerge slightly inclined relative to the east–west (E-W) direction \citep[see \eg ][]{Karak17,Caligari95,Fisher95,Fan94,D'Silva93}. This tendency to have the leading polarity of an AR located towards the solar Equator relative to the following polarity 
was first studied by \citet{Hale19} 
and is referred to as Joy's law \citep{vanDriel15}. Observationally, this law implies that the axis joining the centers of the main polarities of an AR forms an angle, called tilt angle, with respect to the E-W direction. 

%{\S\bf --- Why a good estimation of the tilt is important?} \\
The existence of tilt angles in ARs plays a central part in flux-transport dynamo models, as Joy's law is a fundamental ingredient for the formation and evolution of the polar field \citep[see the review by][and references therein]{Wang17}.  Therefore, obtaining a good estimation of tilt angles, their evolution, and spatial variation on the Sun surface 
plays a key role in constraining this kind of dynamo models. 

%{\S\bf --- White-light data bases that compute the tilt}\\
Tilt angles have been derived since long using databases from white-light (WL) photographic observations taken at Mount Wilson Observatory from 1917 to 1985 and Kodaikanal Solar Observatory from 1906–1987 \citep[see \eg ][]{Howard84,Sivaraman93}. The longest existing catalog of sunspots is the Greenwich Photoheliographic Results \citep[1874\,--\,1976, see \eg ][]{Willis13}. 
After 1976, Debrecen Heliophysical Observatory {developed another WL} %continued with the 
catalog. The Debrecen Photoheliographic Data
is compiled using WL full-disk observations taken at Debrecen Observatory and its Gyula Observing Station \citep{Gyori11,Baranyi16}. There are also two extensions of this database that include magnetic field information. The SOHO/MDI Debrecen data (SDD) includes magnetic and continuum images taken by the Solar and Heliospheric Observatory \citep[SOHO:][]{Scherrer95} with the Michelson Doppler Imager (MDI) instrument, while the  SDO/HMI Debrecen Data uses the magnetic and WL images taken by the Helioseismic and Magnetic Imager \citep[HMI:][]{Schou12} instrument onboard the Solar
Dynamics Observatory \citep[SDO:][]{Pesnell12}. 

%{\S\bf --- Tongues and their influence on tilts}\\
The evolution of photospheric magnetograms is the main source of information on the characteristics of sub-photospheric FRs.
In particular, a noticeable feature is the presence of magnetic tongues \citep[see][and references therein]{Poisson16}. They appear as elongations of the main AR polarities and are mainly observed during the emergence of the top part of $\Omega$-shaped flux ropes. They are produced by the line-of-sight (LOS) projection of the azimuthal component of the FR magnetic field. \citet{Lopez-Fuentes00} were the first to report their existence and, progressively, they were noticed in many other examples \citep[see \eg ][]{Luoni11,Mandrini14,Valori15,Yardley16,Vemareddy17,Dacie18,Lopez-Fuentes18}. These elongated features {are also present} %have also appeared 
in MHD simulations of FR emergence \citep{Archontis10,Cheung10,MacTaggart11,Jouve13,Rempel14,Takasao15}. 
The presence of magnetic tongues naturally modifies the photospheric magnetic distribution of flux concentrations and, therefore, tilt-angle measurements done directly on LOS magnetograms. Furthermore, since sunspots and pores are present in the strongest magnetic fields, magnetic tongues are also expected to modify WL images.

%{\S\bf --- Previous papers and CoFFE} \\
In several articles, we have qualitatively and quantitatively investigated the presence and role of magnetic tongues during the emergence of bipolar ARs. \citet{Poisson15b} presented a systematic method, based on the evolution of the photospheric inversion line (PIL), to quantify the influence of magnetic tongues in emerging FRs. The method allowed us to estimate their average twist, assuming that {the emerging magnetic field} can be represented as a uniformly twisted half torus \citep[see also][]{Luoni11}. \citet{Poisson16} studied how the tongues affect the evolution of the magnetic flux distribution of bipolar ARs, extending the analysis to ARs observed along more than a solar cycle. Since it was found that emerging ARs have a wide set of twist profiles, a more sophisticated FR emergence model was developed that considered FR cross-sections with non-uniform twists (both in the radial and azimuthal directions).  

%{\S\bf --- Removing tongues} \\
However, though in these articles it was shown that the presence of tongues has a non-negligible effect in the determination of the tilt of ARs, none of them developed a method to remove this effect from the intrinsic characteristics of emerging FRs. A method, called Core Field Fit Estimator (CoFFE), has just been presented by \citet{Poisson20}. CoFFE succeeds to remove most of the magnetic tongues effect on the computation of the location of the flux-weighted centers (magnetic barycenters) of the polarities and, hence, it allows to obtain an AR tilt-angle that better represents the FR intrinsic tilt. 

%{\S\bf --- In this paper} \\
In this article, we investigate the role of magnetic tongues on the measurements of tilt angles of sunspot groups derived from WL images. To facilitate the reading of this article in the top block of Table~\ref{tab_acronyms} we list the acronyms most used in our text, their meanings, and the databases to which they refer or are applied to; while in its bottom block we enumerate the different tilt-angle names, the method used to compute them, and the data to which they refer or are applied to. In \sect{Data}, we describe the data we use in our tilt-angle computations. Our methods to compute the tilt values using WL images alone and combining them with magnetic field data, as well as a summary of the CoFFE method applied to magnetograms, are described in \sect{Method}. Next, \sect{Tilt_ARS} presents the results of the application of the previous methods to a set of bipolar ARs with different observed photospheric magnetic flux distributions, \ie~from cases in which tongues are not evident to those with clearly elongated tongues and even some examples with more than one bipole present. We compare the results obtained with these different methods and also with those found in the SDD catalog. Finally, in \sect{Conclusions} we summarize our findings and conclude.

%	TABLE 1  ---------------------------------------------
\begin{deluxetable*}{cll}
\label{tab_acronyms}
\tablenum{1}
\tablecaption{Top block: Acronyms, their meanings, and the data sets to which they refer or are applied to. Bottom block: Tilt-angle names, their associated computing methods, and data sets.}
\tablehead{
\colhead{Acronym} & \colhead{Meaning} & \colhead{Refer/applied to} 
           }
% DPD     & Debreceen Photoheligraphic Data & Debrecen WL data processed with Debrecen software \\
\startdata
 SDD     & SOHO/MDI-Debrecen Data          & MDI WL data processed with Debrecen software plus polarity sign\\
 TM      & threshold method               & MDI WL data \\ 
 k-means & grouping algorithm            & MDI WL data to spatially cluster umbrae\\
 MB      & magnetic barycenters            & MDI LOS magnetograms \\
 CoFFE   & Core Field Fit Estimator        & MDI LOS magnetograms \\
%\enddata
\hline
\multicolumn{1}{c}{Tilt-angle names} & \multicolumn{1}{c}{Computed with} & \multicolumn{1}{c}{Applied to} \\
%\tablehead{
%\colhead{Tilt-angle names} & \colhead{Computed with} & \colhead{Applied to} 
%           }
%\startdata
$\phiwl$    & TM + k-means grouping     & MDI WL data \\
$\phiwlm$   & TM + polarity sign grouping     & MDI WL data \\
$\phia$     & magnetic barycenters (apparent tilt) & MDI LOS magnetograms \\
$\phic$     & CoFFE                                & MDI LOS magnetograms \\
\enddata
\end{deluxetable*}

%%%%%%%%%%%%%%%%%%%%%%%%%%%%%%%%%%%%%%%%%%%%%%%%%%%%%%%%%%%%%%%%%%%%%%%%%%%%%%%%%%%%%%
\section{Data Used}
\label{sect_Data}
%%%%%%%%%%%%%%%%%%%%%%%%%%%%%%%%%%%%%%%%%%%%%%%%%%%%%%%%%%%%%%%%%%%%%%%%%%%%%%%%%%%%%%

%{\S\bf --- SOHO/MDI WL and LOS magnetograms} \\
We use continuum intensity images and LOS magnetograms obtained with MDI.
The full-disk WL images are constructed with the combination of five filtergrams with wavelengths around the 
Ni {\sc i} absorption line. These images have a noise per pixel of $0.3\%$.
The LOS magnetograms are constructed onboard SOHO by measuring the { Zeeman effect} in right and left circularly polarized light.
The magnetograms from the 96-minute series, obtained from 5-minute averaged magnetograms, have lower noise level than the 1-hour series (that includes magnetic and WL data) and an error per pixel of $\approx 9$ G \citep{Liu04}.
Both magnetograms and intensity images have a spatial resolution of $1.98''$ and are digitized with the same CCD with a size of $1024\times1024$ pixels. 
We use all the available WL images from the 1-hour and 1-minute data sets closer in time to the magnetograms from the 96-minute cadence data set.

%{\S\bf --- ARs selection and processing LOS magnetograms} \\
As we aim to characterize the tilt angle evolution in emerging ARs, we selected eight ARs for which we see a clear emergence across their transit through the solar disk.
For all the cases we limit the latitudinal and longitudinal range of the selected ARs within $-35\degree$ to $35\degree$ from the disk center to reduce the foreshortening and limb darkening effects \citep{Green03}. 

%{\S\bf --- All data processing} \\
We process the WL images and the magnetograms to construct two sets of data cubes
for each analyzed AR. Using standard solar software tools, we transform the LOS component of the magnetic field to the solar radial direction. 
As we study ARs located near the solar disk center, the latter approximation produces no significant effect on the resulting magnetic flux density \citep{Green03}.
Next, we rotate the set of magnetograms and WL images to the time when the AR was located at the central meridian. 
This procedure corrects the solar differential rotation using the coefficients derived by \citet{Howard90}. 
Next, we select a sub region which encompasses the AR.  Any image presenting evidence of wrong pixels and/or corrupted data are removed from the set.

%{\S\bf --- WL images processing} \\
In order to detect the umbra regions, we apply a few processing tools from the OpenCV Python 3 package to the WL images.  First we rescale the continuum intensity levels of all the WL images corresponding to the evolution of an AR using the global maximum and minimum of the set. 
Then, we convert the pixel intensities to an unsigned 8-bit integer number, this fixes the number of the intensity levels of the image to $255$. This conversion is in line with previous studies, including the method used with SDD \citep{Gyori98}. Pattern recognition algorithms, including the one used here to detect the umbra, also include this conversion to improve the algorithm performance.
We increase the image contrast in $10\%$ and reduce the brightness in $50\%$ to desaturate the intensity observed in the photosphere.
Finally we apply a 2D filter to emphasize the differences in adjacent pixel values.
This filter performs a linear convolution of the image with a $3\times3$ matrix, or kernel, chosen to increase the image sharpness and, therefore, facilitate the detection of the edges.

%{\S\bf --- SDD catalog for tilt angles} \\
We compare our tilt-angle values deduced from WL umbra detection with those reported in SDD. SDD has free access to ftp data request and an online catalog with sunspot-group information \citep{Gyori10}.
The sunspot groups in this catalog are labeled with the same number as the one assigned by the National Oceanic and Atmospheric Administration (NOAA) to ARs. The catalog combines the image processing algorithms, sunspot detection, and area measurements developed earlier for the Debrecen Photoheliographic Data catalog \citep{Gyori98, Gyori10, Baranyi16}. These techniques are also applied to MDI magnetograms (see \sect{Introduction}), therefore SDD also includes the information on the magnetic polarity signs. 

%%%%%%%%%%%%%%%%%%%%%%%%%%%%%%%%%%%%%%%%%%%
%--------------- METHOD -------------------
%%%%%%%%%%%%%%%%%%%%%%%%%%%%%%%%%%%%%%%%%%%%

\section{Tilt-Angle Estimation Methods} 
%\section{Tilt-angle estimation methods} 
\label{sect_Method}

%%{\S\bf --- Intro sect} \\  
%Among the different data used to compute the tilt angle of ARs, we can distinguish the ones that uses only white light images of sunspots and pores from the ones which relies on the information provided by the magnetic field.  The ARs properties detected with each of these observables (white-light images and magnetograms) can be significantly different. In this section we describe these methods and we summarize their main advantages and disadvantages. 
    
%------------------------------------------------------------
\subsection{Tilt Angle from WL Images} 
%\subsection{Tilt angle from WL images} 
\label{sect_Mod_WL}

%{\S\bf --- Methods to find the umbra} \\
Methods to compute the tilt angle from continuum images start with the identification of the umbra areas within a sunspot group.
These methods can be separated in two groups. The first group corresponds to the threshold methods (TMs), which are based on the selection of a cut-off value for the image intensity levels \citep{Chapman84, Steinegger96}.
The second group are border methods, which use a gradient map to identify the abrupt changes of the image intensity between the umbra-penumbra interface.
The method described by \citet{Gyori98} is an example of the latter group and is the one used on SDD to automatically register the information of umbra areas of sunspot groups.

%{\S\bf --- Detecting umbra/penumbra} \\ 
Tilt angles can be determined from WL images using only the sunspot umbrae or including their penumbrae. The penumbra is in general easier to detect than the umbra at earlier stages of an AR emergence. However, its detection can be affected by the presence of dark penumbral filaments, granular local minima, and/or background magnetic field remnants, which can produce dark features around pores.
Then, tilt angles obtained from area-weighted penumbra centers are frequently strongly affected by these extra features. To avoid determining erroneous tilts, we only consider the values obtained using umbra areas from images processed as summarized in \sect{Data}. In this way tilt values are less noisy, though we have more data gaps at the beginning of the emergence.  

%{\S\bf --- Proximity grouping algorithm} \\ 
Many of the past sunspot records have no magnetic polarity information \citep[see, {\eg}][]{Howard91h}; therefore, it is necessary to use a proximity-based algorithm first to isolate a sunspot group, and then to identify the leading and the following spots or polarities of an assumed bipolar AR.
To do so an area-weighted umbra center of the group is computed, and then the leading (following) portion of the group is assigned to the spots located to the solar west (east) of this center. 

%{\S\bf --- k-means grouping} \\
We use a similar procedure based on the k-means clustering algorithm \citep{macqueen67} to explore the consistency between the different grouping procedures.
This iterative procedure requires the input of the number of  
groups, $k$.  In our case $k=2$, one corresponding to the leading polarity and one to the following one.
Then, each of the separated umbrae are associated to one of these groups.
The routine computes the distance between the center of each umbra to assigned group centers. 
Initially, the group centers are located at random positions within the image; then, the procedure defines new group centers and/or new associations until the global mean distance of each umbra to each group center reaches a minimum.
In other words, the routine seeks to minimize the functional defined as the distance between the umbra centers and the group centers. 
Once an optimal grouping is achieved, we define the group located to the solar west as the leading polarity and the one at the solar east as the following one.

%{\S\bf --- magnetic field grouping} \\
SOHO/MDI magnetograms allow us to use the magnetic field information to separate the leading and the following umbrae.
Using the magnetic-field sign grouping helps us understand the limitations and errors of the methods described above.
In particular, it can identify inconsistencies between the different catalogs due to a wrong assignment of umbrae to the leading or following group; this can result in tilt values computed from sunspots having the same polarity sign \citep{Baranyi15}.

%{\S\bf --- WL tilt angles, definitions} \\
The umbra areas and their polarity sign information let us derive different estimations of the tilt angle. 
The tilt angle is obtained as the acute angle formed between the east-west direction and the line that joins the umbra area-weighted centers of the leading and following polarities. We determine two different tilt angles from the umbra areas, depending on the grouping algorithm.  We define the tilt angle $\phiwl$ derived from the proximity algorithm (k-means grouping) and $\phiwlm$ as the tilt derived considering the magnetic field sign of the umbrae. 
An example is shown in \fig{example}a with the umbra detection done on an MDI WL image corresponding to AR 9906 observed on 2002-04-14.

%{\S\bf --- Way to plot the data} \\
From now on, the data of all ARs are plotted with the same drawing convention (see the caption of \fig{example}).  For each AR the same subregion is shown for WL images and magnetograms (figure panels and associated movies). The spatial coordinates are relative to the bottom left corner, with the X coordinate growing towards the solar west and Y towards the solar north.  When the leading polarity is closer to the equator than the following one, as it is the case for most ARs (Joy's law), we define tilt angle as positive.

%------------------------------------------------------------
\subsection{Tilt Angles from Magnetic Barycenters and Tongues} 
%\subsection{Tilt angles from magnetic barycenters and tongues} 
\label{sect_Mod_MAG}

%{\S\bf --- Computing the magnetic barycenters} \\
LOS magnetograms allow us to study the evolution of AR tilt-angles.
The tilt angle is in general derived from LOS magnetograms using the magnetic barycenters \citep[see][]{Lopez-Fuentes00}.  
Then, as with the WL area-weighted centers, we define the apparent tilt angle, $\phia$, as the acute angle formed between the E-W direction and the segment that joins the barycenters. We call the tilt values derived in this way the magnetic barycenters (MB) method.

%{\S\bf --- Effect of the magnetic tongues} \\
However, the value of $\phia$ is not an exact estimation of the intrinsic tilt angle of the FR that forms the AR \citep{Poisson20}.  As summarized in \sect{Introduction}, the intrinsic FR tilt-angle is modified by the magnetic tongues present during the AR emergence. 
Indeed, the departure of $\phia$ from the intrinsic tilt 
can be significantly larger than the mean dispersion reported in most of Joy's law studies \citep[\eg ][]{Wang15}. 

\begin{figure}[!ht]
\begin{center}
\includegraphics[width=.45\textwidth]{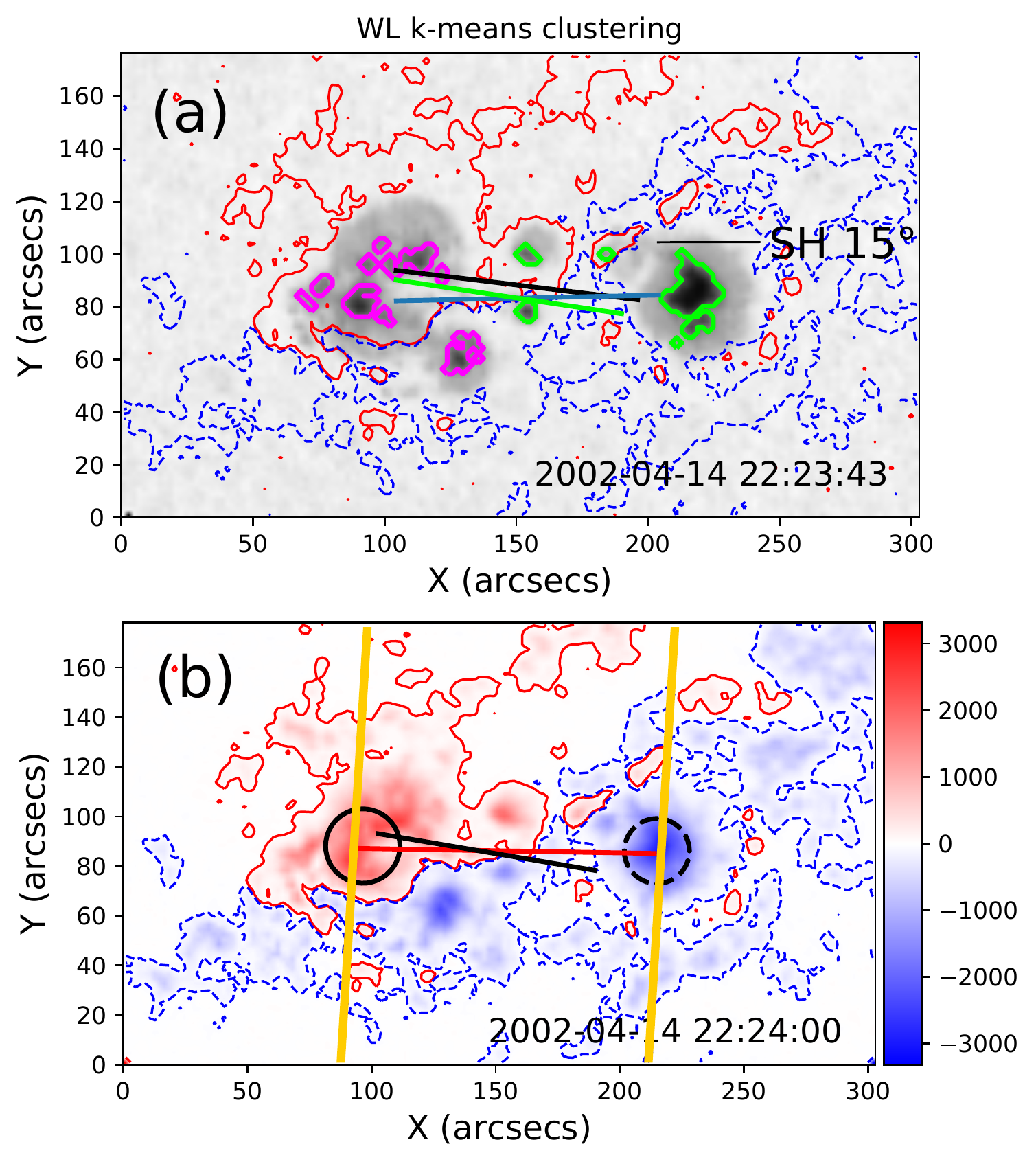}
\caption{SOHO/MDI observations for the southern hemisphere {(SH)} AR 9906: (a) WL image, (b) LOS magnetogram.
   In (a) the green and magenta contours show the umbra areas of the leading and following sunspots, respectively. 
The separation of both sunspot groups is done using the k-means algorithm.
The red (blue) contour corresponds to the positive (negative) magnetic field with a strength of $50$ ($-50$) G.
The blue and green segments indicate the inclination of the bipole from which the values of $\phiwl$ and $\phiwlm$ are obtained, respectively (as defined in \sect{Mod_WL}).
The black segment in both panels corresponds to the bipole inclination computed using the magnetic barycenters, $\phia$ (see \sect{Mod_MAG}). 
  (b) The red- and blue-shaded areas represent the positive and the negative LOS magnetic field component. 
The black circular contours are drawn at the half-maximum height of the CoFFE Gaussian fit for each polarity after convergence, using $p=0$. {These Gaussian fits define the core flux of both magnetic polarities.} 
The red segment shows the tilt of the AR obtained from the core flux centers, $\phic$ (see \sect{CoFFE}).
The yellow lines mark the region in which the magnetic flux of the tongues is removed in the CoFFE iterative procedure. 
Movies of this AR evolution are available as additional material (9906\_WL.mp4 and 9906\_CoFFE.mp4). { From now on, in panels showing WL images and LOS magnetograms, dates in the bottom right corner are indicated in the format year-month-day followed by the time in UT.}
}
 \label{fig_example}
\end{center} 
\end{figure}

%{\S\bf --- Tongues in magnetogram} \\
To illustrate the morphology of magnetic tongues and help us understanding their influence on tilt-angle measurements, we select AR 9906 that has well-developed magnetic tongues (\fig{example}). 
Magnetic tongues are observed in LOS magnetograms, such as the one shown in \fig{example}b, where the red- and blue-shaded areas indicate the positive and negative magnetic polarities, respectively, and where magnetic isocontours of $\pm 50$ G are added with the same color convention. Magnetic tongues are extensions of the leading and following magnetic polarities towards the center of the AR. In this example, the positive polarity (red) extends northward in the direction of the negative polarity (blue), while the negative one has a similar southward extension towards the positive.   
This pair of elongations are recurrently observed in emerging ARs and are interpreted as due to the emergence of a twisted FR \citep[][and references there in]{Poisson20}. Their presence naturally modifies the location of the magnetic barycenters.

%	FIGURE 1  ---------------------------------------------

%--------------------------------------------------------------
\subsection{Tilt Angles from the Core Field Fit Estimator (CoFFE)}
%\subsection{Tilt angles from the Core Field Fit Estimator (CoFFE)}
\label{sect_CoFFE}

%{\S\bf --- General idea of the method} \\
 The CoFFE method is based on the identification of two different magnetic flux components that produce the LOS magnetic field distribution observed in emerging ARs \citep{Poisson20}. These components are noted as core and tongue fluxes.
We associate the core flux to the flux of the axial field of a toroidal FR during its emergence \citep{Poisson20}. The tongue flux is the magnetic flux in the elongations of the magnetic polarities, due to the FR azimuthal field component, as previously described.
The core flux is modeled using a 2D Gaussian. Its fit to the corresponding field distribution in the magnetogram provides the core center of each polarity. Then, the tilt angle, $\phic$, is computed using the core centers, as done when using the magnetic barycenters.

%{\S\bf --- CoFFE iterations $n_{\rm it}$} \\
More precisely, the CoFFE method performs simultaneously a fit of the field distribution of each polarity with a Gaussian to isolate the core field and 
removes the tongue component of the field distribution. 
To do so an iterative procedure is designed.
An initial fit to each polarity flux provides a rough estimation of the core centers. Then, an exclusion region is defined.  This region is delimited with two lines perpendicular to the line joining the core centers and crossing each of them  
(see yellow lines in \fig{example}b).
This region is typically located where the tongue contribution is dominant over the core.
So in the first iteration, a new fit of the core is done removing the points in the exclusion region from the fitting procedure.
Finally, iterations are performed until a convergence criteria over $\phic$ is fulfilled.
In order to improve the performance of the method a smaller or larger exclusion region can be defined  \citep{Poisson20}.  For our aim, in this work, it is enough to use the just described basic CoFFE method for the studied ARs.

%{\S\bf --- CoFFE analysis of the AR evolution} \\
To ensure a good approximation of the core region, we start our computation with the magnetogram that is closer to the AR maximum flux. At this time, we expect that the core flux be stronger than the one of the tongues, and therefore, easier to identify and constrain. Once the iteration procedure is completed for this magnetogram we use the obtained Gaussian parameters as an initial guess for the fit in the previous magnetogram towards the beginning of the emergence. In this way a progressive procedure is used in which the core parameters computed at time step $i + 1$ are used to initiate the computation at time step $i$.

%{\S\bf --- Fig 1b} \\
 An example of the application of CoFFE to a LOS magnetogram of AR 9906 is shown in \fig{example}b.
The black circles correspond to the isocontours of the Gaussian function fitted to the core flux of each polarity.
The level of these contours is set to $50\%$ of their respective Gaussian maximum value.
The red segment connecting the center of the positive and negative core regions corresponds to the inclination of the bipole computed with CoFFE, from which we derive the tilt $\phic$. 
The black segment corresponds to the tilt $\phia$ computed from the magnetic barycenters, {or apparent tilt}. This segment shows the shift of the magnetic barycenters towards the center of the AR due to the presence of strong magnetic tongues. 

%{\S\bf --- Summary of CoFFE results/performances} \\
 A series of tests on FR models and ARs have shown that CoFFE provides a better estimation of the tilt angle since it removes efficiently the effect of the magnetic tongues \citep{Poisson20}.
The correction achieved with $\phic$ requires just a little more computational effort than the previously described methods. 
Finally, the removal of the effect of the tongues allows us to expand the determination of AR tilts to the early stages of their emergences, since magnetic tongues are typically stronger at the beginning of the emergence \citep[dominance of the azimuthal field component at the top of FR,][]{Poisson16}.

%	FIGURE 2  ---------------------------------------------
\begin{figure}[!ht]
\begin{center}
\includegraphics[width=.45\textwidth]{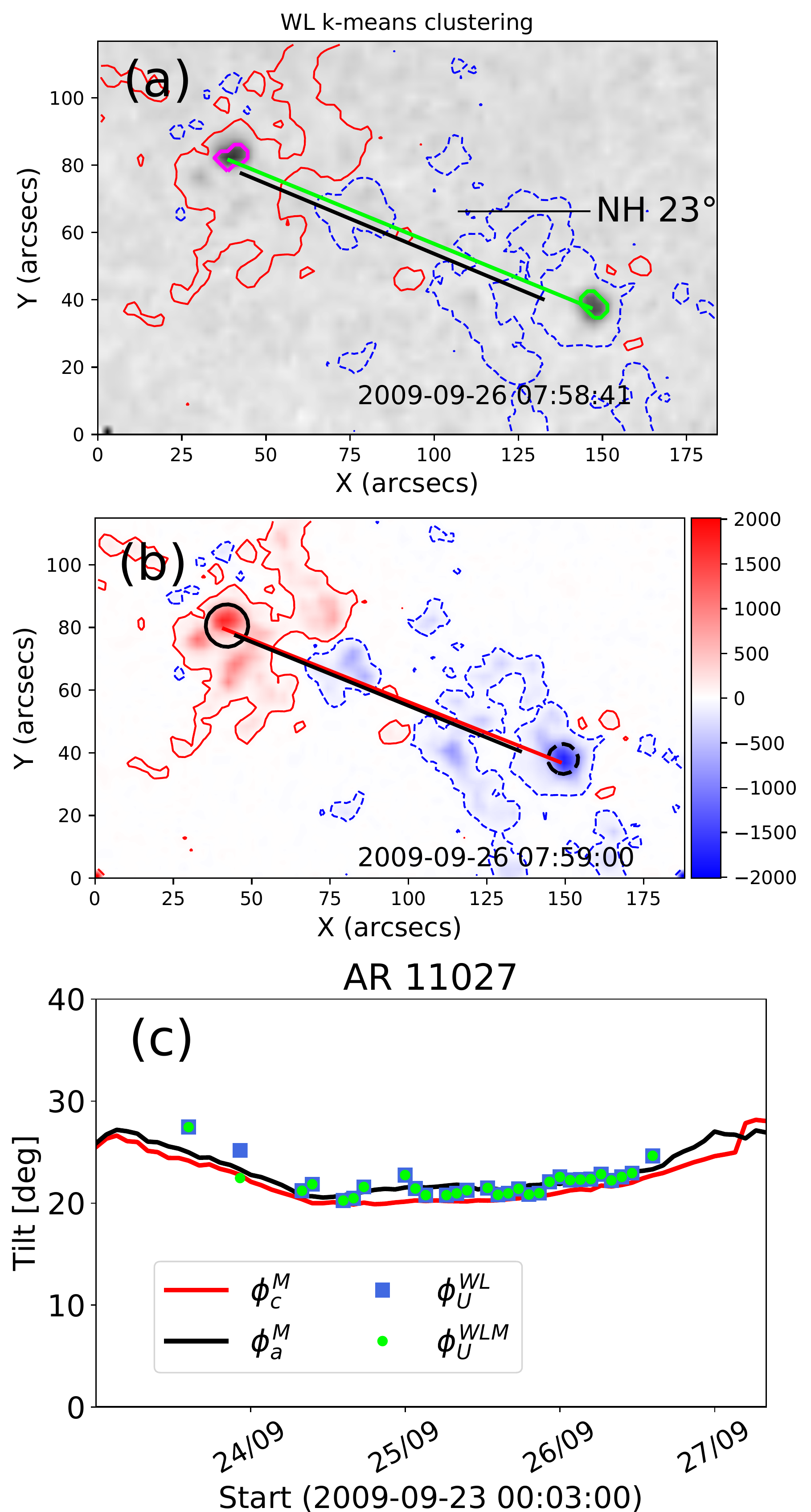} \\
\caption{SOHO/MDI observations of the northern hemisphere {(NH)} AR 11027: (a) WL image, (b) LOS magnetogram. 
The drawing convention is the same as in \fig{example}a,b.
(c) Evolution of the tilt angles along the emergence of AR 11027. 
The black continuous line corresponds to the tilt angles computed from LOS magnetograms and the magnetic barycenters method described in \sect{Mod_MAG}. The red line shows the title-angle values computed with CoFFE (\sect{CoFFE}).
The blue squares correspond to the tilt angles obtained from WL images using the k-means clustering method, 
while the green dots {represent} the tilt angles computed using WL images and magnetic field grouping (\sect{Mod_WL}). { In this and panels with similar information the bottom label indicates the date in the format year-month-day followed by the time in UT.}
{Movies of this AR evolution are available as additional material (11027\_WL.mp4 and 11027\_CoFFE.mp4)}.
}
 \label{fig_11027_tilt}
\end{center} 
\end{figure}
%--------------------------------------------------------------  

%	FIGURE 3  ---------------------------------------------
\begin{figure}[!ht]
\begin{center}
\includegraphics[width=.45\textwidth]{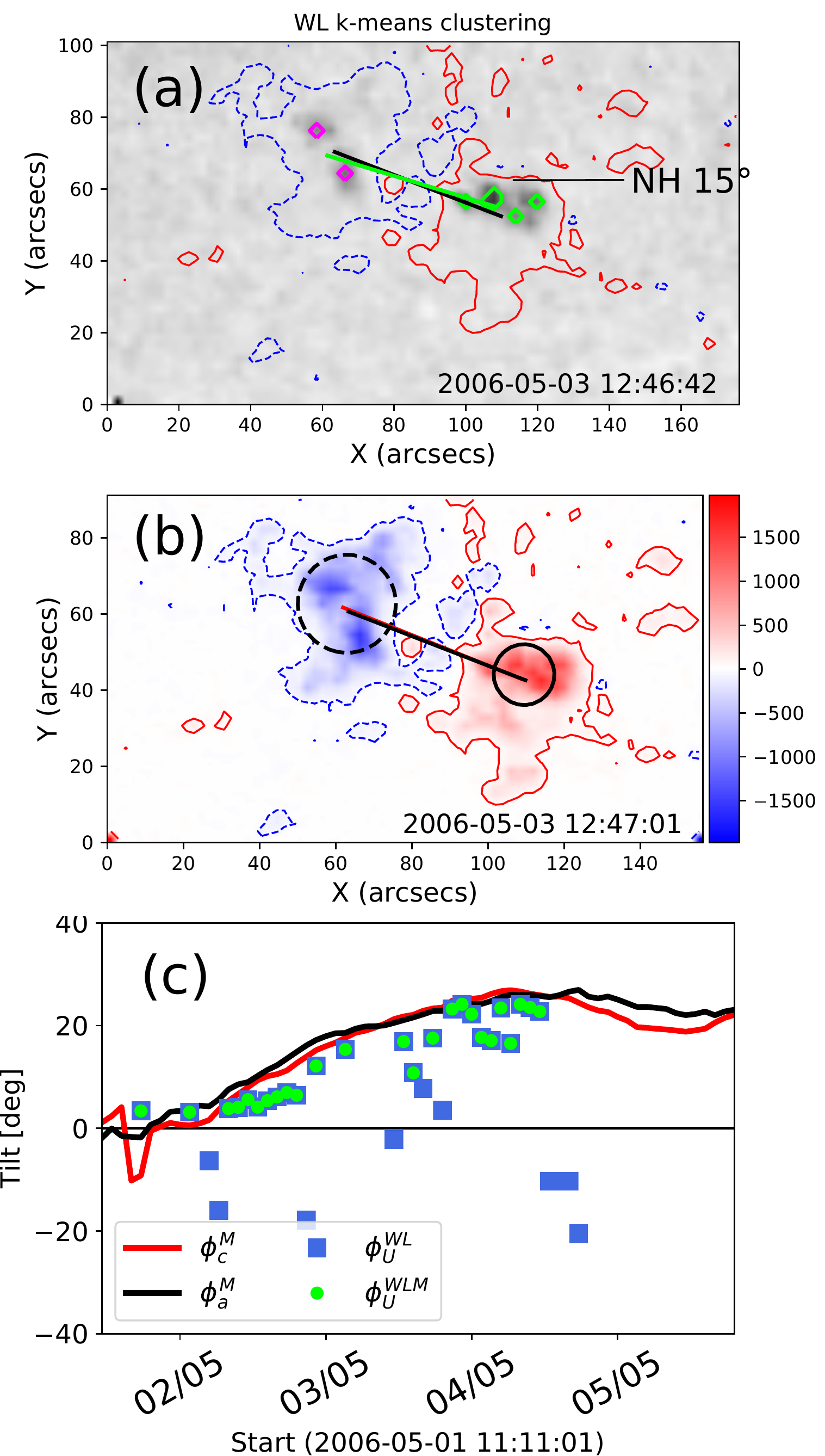} 
\caption{SOHO/MDI observations of the northern hemisphere {(NH)} AR 10879: (a) WL image, (b) LOS magnetogram.
(c) Evolution of the tilt angles along the emergence of AR 10879. 
The drawing convention is the same as in \figs{example}{11027_tilt}. 
Movies of this AR evolution are available as additional material (10879\_WL.mp4 and 10879\_CoFFE.mp4).
}
 \label{fig_10879_tilt}
\end{center} 
\end{figure}
%--------------------------------------------------------------

%%%%%%%%%%%%%%%%%%%%%%%%%%%%%%%%%%%%%%%%%%%
%--------------- RESULTS -------------------
%%%%%%%%%%%%%%%%%%%%%%%%%%%%%%%%%%%%%%%%%%%%

\section{Comparing the Tilt Angle Estimation Methods}
%\section{Comparing the tilt angle estimation methods}
 \label{sect_Tilt_ARS} 

%{\S\bf --- The variety of cases we use} \\
To illustrate the effect of magnetic tongues on the estimation of tilt angles computed using LOS magnetograms, both $\phia$ and $\phic$, and WL observations, both $\phiwl$ and $\phiwlm$, we select a series of ARs. 
In \sect{Bipolar} we start analyzing the emergence of bipolar ARs in which tongues are small and weak (\sect{Weak_tongues}) and continue with ARs that have extended and strong tongues all along their emergence phase (\sect{Strong_tongues}). Next, in \sect{SumBi}, we  summarize the main characteristics and results obtained for bipolar ARs.
Finally, we deal with two ARs in which the evolution of the main bipole is accompanied by the emergence of secondary bipoles (\sect{Multipolar}). 
This variety of examples lets us explore the performance of the methods described in \sect{Method} for the computation of AR tilts, as well as their validity and consistency. 

%------------------------------------------------------------
\subsection{Bipolar ARs}
\label{sect_Bipolar}

%---------------------------------------------
\subsubsection{ARs with Small and Weak Tongues}
%\subsubsection{ARs with small and weak tongues}
\label{sect_Weak_tongues} 

%{\S\bf --- Generalities of this section} \\
In this section, we show two examples of ARs with small and weak (low magnetic field intensity) tongues, AR 11027 and AR 10879. Both ARs emerge in the northern hemisphere in a low background field region. In these cases, tongues are visible only in the first days of the emergence and, sometimes, they are clear in only one of the two main polarities.

%{\S\bf --- Figs. 2a-b and 3a-b - Segments and movies} \\
Panels (a) and (b) in \figs{11027_tilt}{10879_tilt} show snapshots of the evolution of AR 11027 and AR 10879, respectively, as seen in 
WL images (panels a) and MDI LOS magnetograms (panels b). 
The green segments in panels {(a)} join the location of the leading and the following umbra centers computed using the magnetic field polarity information {for clustering}.
In both figures the blue segments that join the umbra centers, computed
using the threshold method and k-means grouping, completely
agree with the green segments {(which mask them)}.
The black segments in panels (a) and (b) of \figs{11027_tilt}{10879_tilt} join the magnetic barycenters, while the red segments in panels (b) connect the polarities core centers computed using CoFFE.
The evolution of these segments as the ARs emerge can be followed in the WL {(11027\_WL.mp4 and 10879\_WL.mp4)} and MDI LOS magnetogram {(11027\_CoFFE.mp4 and 10879\_CoFFE.mp4)} movies.

%{\S\bf --- Panels c of Figs. 2 and 3 - Magnetograms} \\
Panels (c) in \figs{11027_tilt}{10879_tilt} illustrate the evolution of the four tilt-angle measurements described in \sect{Method} and 
Table~\ref{tab_acronyms}. For values derived using LOS magnetograms, the coincidence between $\phia$ and $\phic$ is evident in the associated movies, i.e. both black and red continuous lines globally follow the same behavior. 
Furthermore, the tilt values are in general positive which agrees with Joy's law.
There are only a few negative values of $\phic$  in the early emergence of AR 10879 (\fig{10879_tilt}c) that are due to the disperse core flux {(its center cannot be clearly determined when fitting the Gaussian function). $\phia$ is more stable for this early emergence phase ($\phia \approx 0$). Finally, AR 10879 is an example where an intrinsic clockwise rotation of the bipole is well identified.

%{\S\bf --- Panels c of Figs. 2 and 3 - White light} \\
Concerning WL tilt-angle measurements for AR 11027, they closely agree (see the blue squares and green dots in  \fig{11027_tilt}c). Generally speaking, the four tilt values remain being close during all the emergence phase.
 The same is true for AR 10879, except for several WL tilt values (blue squares) as we explain in the next paragraph. 

%{\S\bf --- Panel c of Figs. 3 - Unipolar cases} \\
\Fig{10879_tilt}c shows several values of $\phiwl$ that are not accompanied by the corresponding ones of $\phiwlm$. For these cases all umbra centers belong to same polarity producing fake bipolar identifications from unipolar configurations. The disperse flux of the following polarity forms weak umbrae which are not detected during a few time intervals of the AR evolution, while umbrae are always present in the leading polarity. 
This implies that WL tilt determinations should be limited to those ARs in which the magnetic flux density is large enough to form umbrae in both polarities. This introduces a strong bias in a large number of tilt-angle measurements that only use WL data \citep{Baranyi15}. 

%	FIGURE 4  ---------------------------------------------
\begin{figure*}[!ht]
\begin{center}
\includegraphics[width=0.9\textwidth]{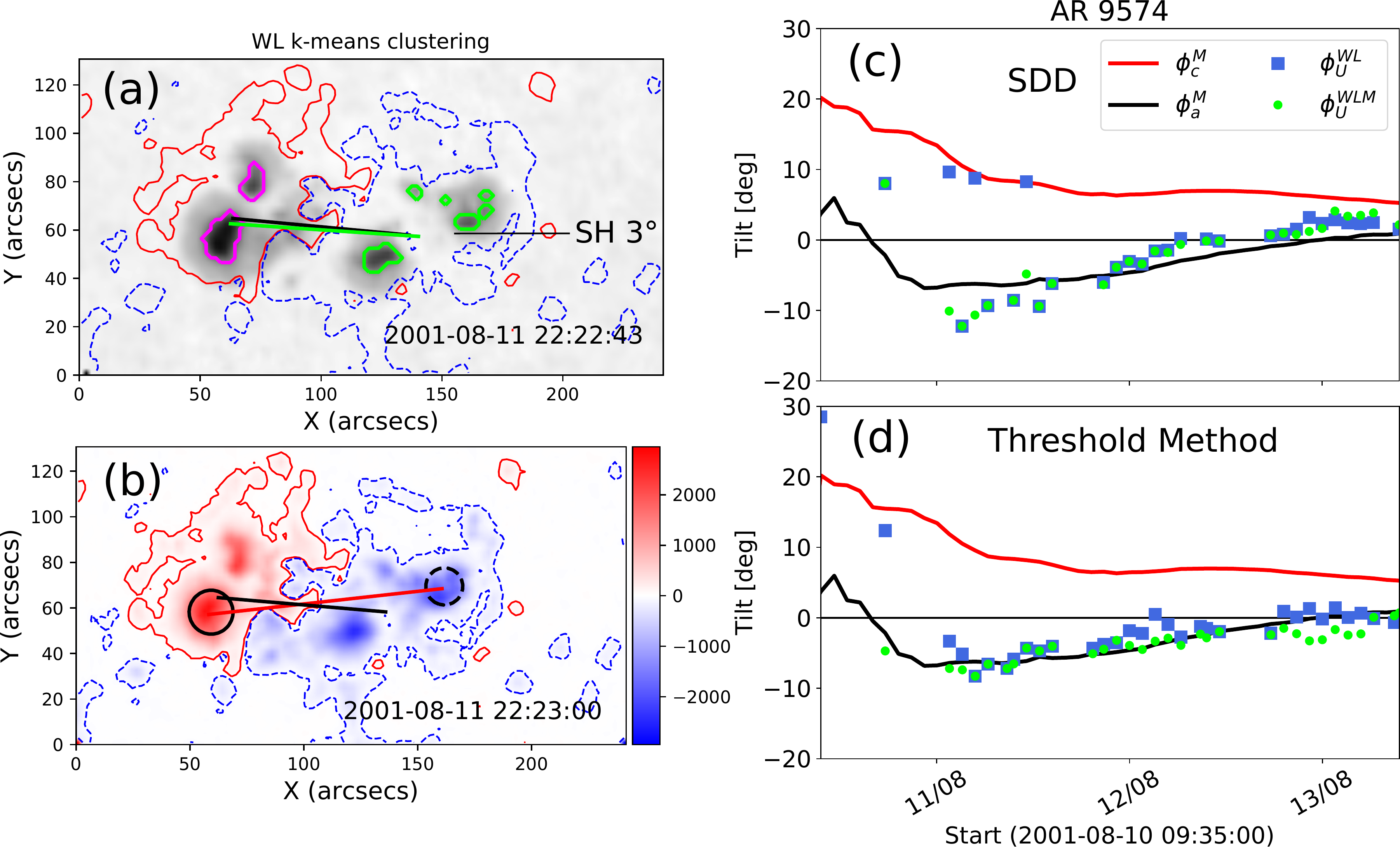} 
\caption{SOHO/MDI observations of the southern hemisphere {(SH)} AR 9574: (a) white light image, (b) LOS magnetogram. 
The drawing convention is the same as in \fig{example}a,b.
(c)-(d) Evolution of the tilt angles along the emergence of AR 9574.
 (c) Comparison between $\phiwl$ and $\phiwlm$ obtained from the SDD catalog, and (d) from the umbra detection with a threshold method. 
In (c) the blue squares correspond to the tilt values obtained from the SDD proximity grouping method and in (b) using k-means clustering (\sect{Mod_WL}). The green dots in both panels show the tilt angles $\phiwlm$ computed including the magnetic field sign information to both grouping algorithms.
The black and red continuous lines in both panels have the same meaning as those in \fig{11027_tilt}c.
Movies of this AR evolution are available as additional material (9574\_WL.mp4 and 9574\_CoFFE.mp4).
}
 \label{fig_9574_tilt}
\end{center} 
\end{figure*}
%-------------------------------------------------------------
 
%	FIGURE 5  ---------------------------------------------
\begin{figure}[!ht]
\begin{center}
\includegraphics[width=.45\textwidth]{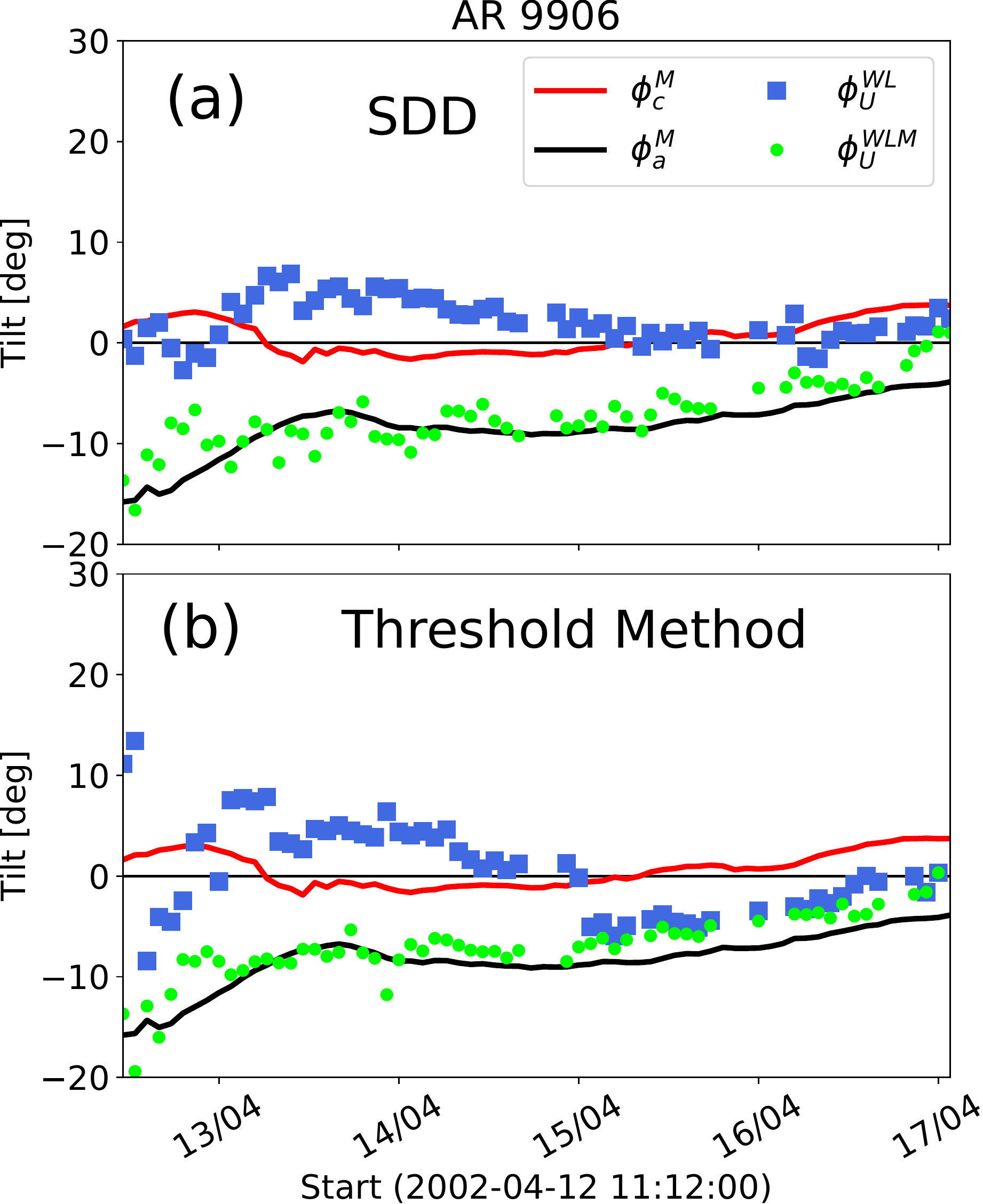}
\caption{Evolution of the tilt angle along the emergence of the southern hemisphere AR 9906. 
Comparison between the $\phiwl$ and $\phiwlm$ using the data from the SDD catalog in (a) and 
the umbra detection using a threshold method in (b). 
The black and red continuous lines in both panels have the same meaning as those in \fig{11027_tilt}c.
In (a) the blue squares correspond to the tilt values obtained from the SDD proximity grouping method and in (b) using k-means clustering (\sect{Mod_WL}). The green dots in both panels show the tilt angles $\phiwlm$ computed including the magnetic field sign information to both grouping algorithms. 
{Movies of this AR evolution are available as additional material (11027\_WL.mp4 and 11027\_CoFFE.mp4).}
}
 \label{fig_9906_tilt}
\end{center} 
\end{figure}
%--------------------------------------------------------------

%---------------------------------------------
\subsubsection{ARs with Extended and Strong Tongues}
%\subsubsection{ARs with extended and strong tongues}
\label{sect_Strong_tongues}

%{\S\bf --- Generalities of this section}\\
We select AR 9906 and AR 9574 to illustrate the influence of extended and strong (high magnetic field intensity) tongues on tilt-angle measurements. Both ARs emerge in the southern solar hemisphere and have tongues all along their emergence, even when reaching their maximum magnetic flux. In both cases the most extended and strong tongue is the one of the leading polarity. 

%{\S\bf --- Parallel between Fig.1a-b and \fig{9574_tilt}a,b - Segments and movies}\\
\Fig{example}a,b and \fig{9574_tilt}a,b show snapshots of the evolution of AR 9906 and AR 9574, respectively, as seen in
MDI WL (panels a) and LOS magnetograms (panels b). Notice that in both cases tongues are so strong that umbrae are present in WL images at these elongated regions. The blue, green, black, and red segments in \fig{9574_tilt}a,b are equivalent to those defined in \fig{example}a,b (see also \sect{Weak_tongues}). The evolution of these segments as the ARs emerge can be followed in the WL ({9906\_WL.mp4} and {9574\_WL.mp4}) and MDI LOS magnetogram ({9906\_CoFFE.mp4} and {9574\_CoFFE.mp4}) movies.

%{\S\bf --- Description of \fig{9906_tilt}a,b AR 9906 - Magnetograms}\\
The black and red continuous lines in \fig{9906_tilt}a,b show the evolution of the apparent tilt angle, $\phia$ {(black curve)}, and that derived using the core flux centers, $\phic$ {(red curve) for AR 9906 (shown in \fig{example})}. Conversely, to what is observed in the case of ARs with small and weak tongues, these values do not agree. On one hand, $\phia$ stays always negative contrary to what is expected from Joy' s law, and on the other hand a counter-clockwise rotation of around 10$\degree$ is present all along the AR emergence. These two behaviors are induced by the presence of the extended and strong tongues and, as shown by the evolution of the red curve, they disappear when $\phic$ is computed using the CoFFE method. The values of $\phic$ stay close to 0$\degree$ and their variation do not indicate any clear bipole rotation along the AR emergence.  

%{\S\bf --- Description of \fig{9574_tilt}c,d AR 9574 - Magnetograms}\\
In the case of AR 9574, the value of $\phia$ {(black curve)} is positive {during a short time} at the beginning of the emergence 
in agreement with Joy's law {(\fig{9574_tilt}c,d)}.  It then turns to be negative changing by {about 14$\degree$,} %less than 15$\degree$, 
implying a clockwise rotation of the bipole forming the AR. This rotation changes to be counter-clockwise  by $\approx$ 10$\degree$ after the first emergence day, returning to 0$\degree$ by the end of the emergence period. This behavior would imply that the AR is formed by a flux rope having first a {FR axis with a negative writhe and later a positive one \citep[see \eg\ ][for the link between tilt rotation and writhe]{Lopez-Fuentes03}.}
However, the values of $\phic$ stay always positive in agreement with Joy's law and the evolution of $\phic$ implies a consistent clockwise rotation by around 15$\degree$. 
{We conclude that} magnetic tongues affect the determination of the tilt angle derived from LOS magnetograms changing both its value and the rotation direction of AR 9574. 
  
%{\S\bf --- Why SDD compared to k-means}\\
The examples with extended and strong tongues give us the chance to explore the influence of the grouping algorithms in the case of using only WL observations. {Indeed,} as shown in \fig{example}a and \fig{9574_tilt}a, umbrae are present at tongue locations affecting the way algorithms either based on proximity (k-means) or magnetic field grouping work. 

%{\S\bf --- Description of \fig{9906_tilt}a,b AR 9906 - White light}\\
The blue squares in \fig{9906_tilt}a depict the results derived from the proximity algorithm used by SDD to group umbrae, while the same symbols illustrate the results for the threshold method and k-means grouping in \fig{9906_tilt}b. The grouping done using k-means assigns large umbra areas located on each of the magnetic tongues to the opposite magnetic polarity group, see the southern (northern) umbrae with magenta (green) contours at the center of \fig{example}a.
The grouping algorithm used by SDD also does a similar association.
For both grouping algorithms, the tilt results (blue squares)} are closer to those found with CoFFE, compare with the red continuous line in \fig{9906_tilt}a,b. This means that, for this AR and similar configurations, the grouping algorithms remove efficiently the tongue effect during most of the emergence phase. This removal is less efficient for the k-means grouping than SDD at the beginning and at the end of the emergence, when the blue squares get closer to the green points in \fig{9906_tilt}b. In contrast, the grouping made including the magnetic field sign information (green dots in both panels) gives values which are strongly affected by the tongues and, then, mostly equivalent to the apparent tilt-angle values (black continuous line in both panels).  
 
%{\S\bf --- Description of \fig{9574_tilt}c,d AR 9574 - White light}\\
The tilt-angle correction, found for AR { 9574} using only the grouping procedures, cannot be generalized to other cases, as shown by \fig{9574_tilt}c,d.
The umbrae at the locations of magnetic tongues are here properly assigned to their respective magnetic polarity sign by both the SDD proximity algorithm and the k-means grouping.
So, there is roughly no difference between the estimations of the tilt without ($\phiwl$) and with ($\phiwlm$) magnetic information (see blue squares and green dots in \fig{9574_tilt}c,d), except for a few values. Furthermore, as the umbra areas located at the tongues are as large as the ones in the core, the tilt angles derived from WL images follow the behavior of the apparent tilt values (black continuous curve in \fig{9574_tilt}c,d) while they significantly depart from the tilt values estimated with CoFFE (red continuous curve in \fig{9574_tilt}c,d).

%%{\S\bf --- Summary}\\
%The results for both ARs, with extended and strong tongues, indicate that the CoFFE method successfully removes the influence of magnetic tongues in the measurements of tilt angles derived from LOS magnetograms. However, one should bear in mind that for the CoFFE method to be applicable and useful, one needs to have the emergence evolution as complete as possible in order to identify well the FR core at some point and proceed backwards in time with the analysis. As concerns WL observations, the examples in this section show that magnetic tongues can have a counterpart in WL. This implies that tongues can affect tilt-angle estimations derived from WL data in a similar way as they affect the tilt estimations from LOS magnetograms.

%---------------------------------------------
\subsubsection{Summary of Bipolar ARs Characteristics} 
%\subsubsection{Summary of bipolar ARs characteristics} 
\label{sect_SumBi}

%{\S\bf --- ARs with small and weak tongues} \\
ARs with small and weak tongues are the easiest to analyze, since similar tilt values are expected to be obtained using the four described methods, those derived from LOS magnetograms ($\phia$ and $\phic$) and those from WL images ($\phiwl$ and $\phiwlm$).  
This is illustrated by the global agreement shown in \fig{11027_tilt}c and \fig{10879_tilt}c of the four tilt-angle values.

%{\S\bf --- Unipolar regions} \\
However, even for those simple ARs, tilt measurements using WL data could be incorrect because one of the AR polarities could have no umbra.
This results in WL tilt measurements done only on one magnetic polarity, which has no meaning.  This happens mostly at the beginning of the AR emergence, while it could be also present later on.  This problem should be solved when measuring the tilt using LOS magnetograms.
 
%{\S\bf --- ARs with large tongues: magnetograms and WL}\\
Conversely, tilt measurements could be strongly modified (up to $\approx 20 \degree$) by magnetic tongues if they are extended and strong.  The examples in \sect{Strong_tongues} show that magnetic tongues can have umbrae in WL. This implies that tongues can affect tilt-angle estimations derived from WL data.  Tongues also affect the tilt measurements derived from the computation of the magnetic barycenters.  This leads to false tilts, that can even be in disagreement with the Joy's law. Furthermore, these wrong determinations can as well lead to infer a spurious rotation of the AR bipole, which can even change of direction during the AR emergence.   In contrast, the CoFFE method successfully removes the influence of magnetic tongues in the measurements of tilt angles derived from LOS magnetograms. However, one should bear in mind that for the CoFFE method to be applicable and useful, one needs to have the emergence evolution as complete as possible in order to identify well the FR core at some point and proceed backwards in time with the analysis. 

%{\S\bf --- A variety of tongue morphologies and evolutions}\\
We have illustrated the results just discussed using four ARs with well-defined tongue characteristics (see \sect{Weak_tongues} and \sect{Strong_tongues}).  Still, AR emergences have a broad range of tongue morphologies and evolutions as shown by \citet{Poisson16} in their study of mainly bipolar ARs covering a full solar cycle. To get a glance of this variety, we present in \app{Variety_tongues} the results using two additional ARs. Our analysis shows that the previous results are quite general,  except that the fake or biased deduced tilts could evolve in different ways according to the tongue evolution during the AR emergence. However, CoFFE, in general, gives more stable tilt values and, in particular, eliminates the spurious bipole rotations inferred when using WL data or the barycenter method applied to magnetograms.

%------------------------------------------------------------
\subsection{Multipolar ARs} 
\label{sect_Multipolar}

%{\S\bf --- Multipolar general}\\
In most works {\citep[{\eg}][]{Tlatova18,Illarionov15,Li12,Stenflo12,Tlatov10}} tilt angles are measured using either a single LOS magnetogram or WL image per day, so without studying the evolution of the AR. In these articles all ARs are included despite being monopolar, bipolar or multipolar. In this section, we test the different methods summarized in \sect{Method} and apply them to multipolar ARs for which defining a tilt angle is very difficult. We present two examples of ARs. In the first case, AR 11007, a secondary bipole appears with a reverse polarity sign between the main polarities of the first emerged bipole. In the second case, AR 9748, flux emergence makes the distribution of the flux in the tongues change from displaying a positive twist-like pattern to a negative one.   

\begin{figure}[!ht]
\begin{center}
\includegraphics[width=.45\textwidth]{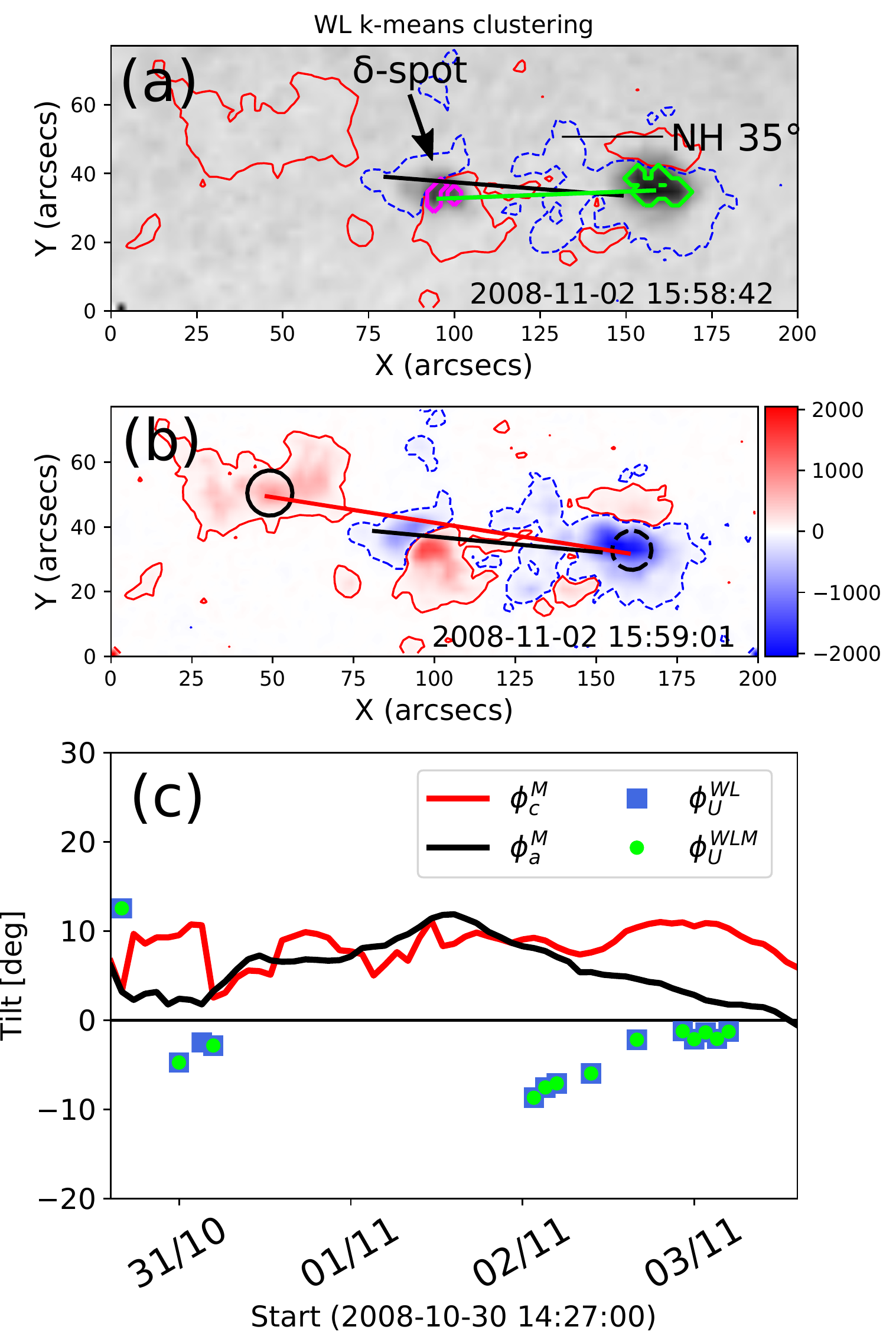} 
\caption{SOHO/MDI observations of the northern hemisphere {(NH)} AR 11007: (a) white light image and (b) LOS magnetogram. 
The arrow in panel (a) points to the secondary emerging bipole. 
(c) Evolution of the tilt angles along the emergence of AR 11007. % Black and red continuous lines, as well as blue squares and green dots, have the same meaning as those in \fig{11027_tilt}c. 
The drawing convention is the same as in \figs{example}{11027_tilt}. Movies of this AR evolution are available as additional material (11007\_WL.mp4 and 11007\_CoFFE.mp4).
}
 \label{fig_11007_tilt}
\end{center} 
\end{figure}

%{\S\bf --- AR 11007 - General} \\ 
Despite the disperse magnetic flux and small concentrated polarities,
AR 11007 has clear magnetic tongues which are visible during the first half of the emergence.
Towards its end, when tongues have almost retracted, a delta group emerges between both polarities
(see \fig{11007_tilt}a,b, the LOS magnetogram movie, 11007\_CoFFE.mp4, and WL movie, 11007\_WL.mp4).  

%	FIGURE 6  ---------------------------------------------
%--------------------------------------------------------------  

%{\S\bf --- AR 11007 - Magnetograms} \\
The values of $\phia$ (black continuous curve in \fig{11007_tilt}c) are strongly affected by the flux in the tongues, as well as by the presence of the central bipole from its early emergence in the late hours of 1 November 2008 (at around 19:15 UT). 
The evolution of $\phia$ shows two successive rotations of the AR, first clockwise and later counter-clockwise. The second rotation is just a spurious effect due to the evolution of the central bipole and is not related either to the presence of tongues or the intrinsic rotation of the main bipole. Next, {close to the beginning of the emergence, some $\phic$ measurements  (red continuous curve in \fig{11007_tilt}c, on 31 October 2008} from $\approx$ 04:45 UT to $\approx$ 12:45 UT) are affected by the stronger flux in the tongue compared to that in the core of the following polarity. This shifts the position of its Gaussian center {and provides lower $\phic$ values}. After this period of time, the CoFFE method  provides a more stable tilt because the exclusion region, defined to remove the tongues, also removes the emerging bipole around the AR center. Tilt values derived from CoFFE agree with what is expected from Joy's law and indicate no clear rotation of the main bipole.     
 
%{\S\bf --- AR 11007 - White light} \\
AR 11007 provides the opportunity to illustrate several problems of tilt-angle measurements using WL data. From the first four measurements (see blue squares and green dots in \fig{11007_tilt}c and the evolution in movie {11007\_WL.mp4}) only the first one corresponds to correct groupings either using k-means or polarity signs. 
The other three measurements have a wrong bipole determination, and even a monopolar region is present for the middle one.
These type of problems were already encountered in \sect{Weak_tongues} and \app{Variety_tongues}.  
   After that, none of the two polarities have detectable umbrae (notice the large gap in the WL data) until the field intensity in the leading polarity is enough to produce umbrae, while this is not the case for the disperse following polarity. This happens at around 2 November at 03:10 UT.  Then, the umbra group center of the preceding negative polarity of the main bipole is falsely associated to the single positive umbra of the new bipole. This happens up to the end of the emergence resulting in wrong estimations of $\phiwl$ and $\phiwlm$ (\fig{11007_tilt}c).

%	FIGURE 7  ---------------------------------------------
\begin{figure*}[!ht]
\begin{center}
\includegraphics[width=.8\textwidth]{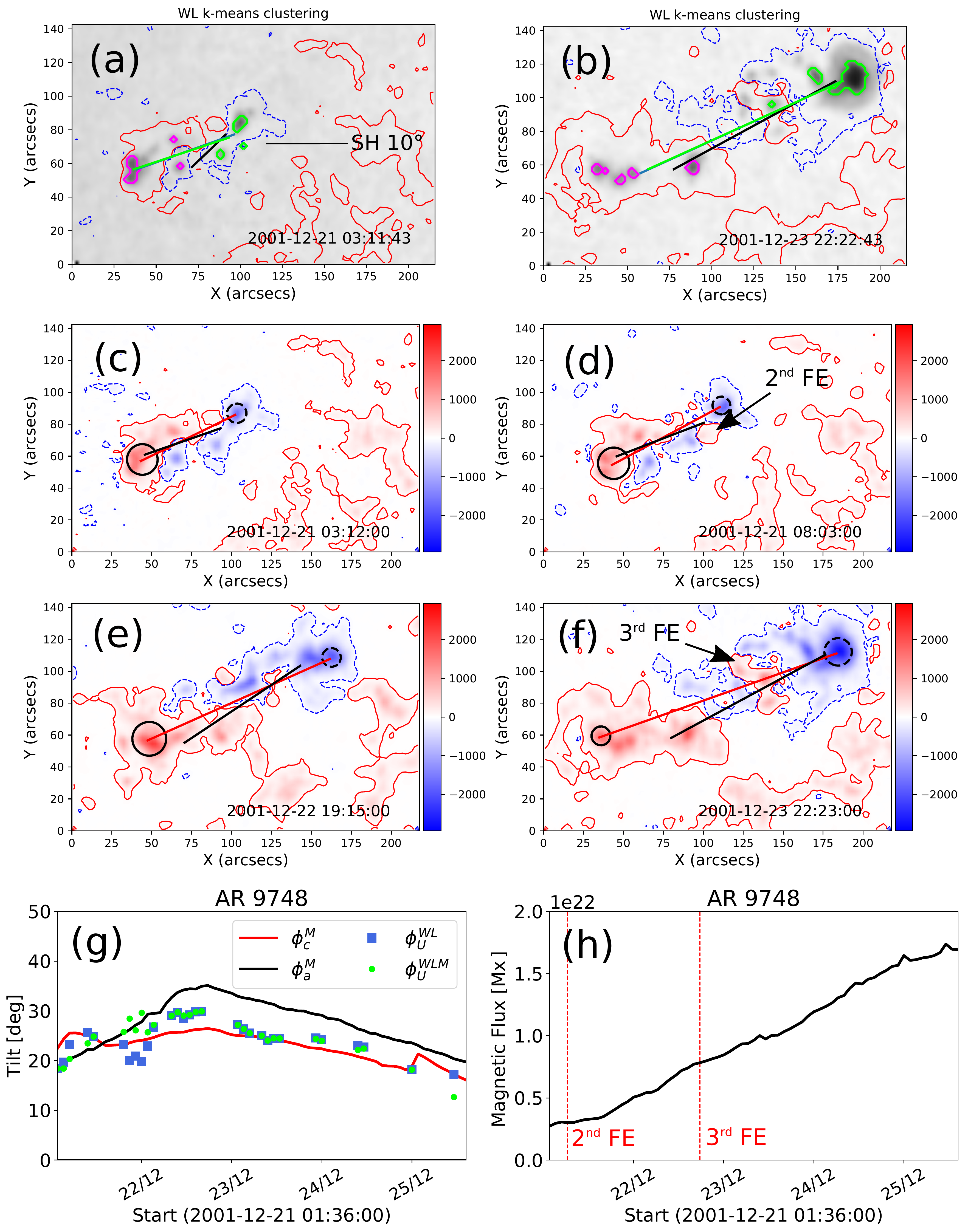}
\caption{SOHO/MDI observations of the southern hemisphere {(SH)} AR 9748: { (a)-(b) white light images, (c) -- (f) LOS magnetograms. 
(g) Evolution of the tilt-angle values along the emergence of AR 9748.
(h) Evolution of the unsigned magnetic flux (see text). The vertical red lines indicate the times when the second and third flux emergences (FEs) are first observed.}
The drawing convention is the same as in \figs{example}{11027_tilt}.
Movies of this AR evolution are available as additional material (9748\_WL.mp4 and 9748\_CoFFE.mp4).
}
 \label{fig_9748_tilt}
\end{center} 
\end{figure*}
%-------------------------------------------------------------- 

%{\S\bf --- AR 9748- Elongated and fragmented tongues}\\
{ Six snapshots of the evolution of AR 9748, two WL images and four MDI magnetograms, are shown in \fig{9748_tilt}.}
The AR is clearly bipolar in its early emergence with elongated and weak tongues. The tongue flux, mainly the one of the preceding polarity, is fragmented in the first stages of the emergence.  By the second half of 21 December, a secondary bipole emerges in between the main one { (see panel d)}. 
At the beginning of the AR emergence the distribution of the flux in the tongues indicates a positively twisted FR; but by mid 22 December, the flux distribution is rather compatible with a negatively twisted FR.   This apparent change in the FR twist is produced by false tongues due to new flux emergence (compare panels c and e of \fig{9748_tilt}). Next, there is a third bipole emergence seen at around 22 December at $\approx$ 21:00 UT, when a positive polarity starts distorting the shape of the negative elongated false tongue (see panel f and the LOS magnetic field movie 9748\_CoFFE.mp4) and a negative polarity appears later to its east. { The flux in the third bipole is lower than in the second one and does not alter the evolution of the total unsigned magnetic flux (positive flux plus absolute value of the negative one divided by 2) shown in panel (h)}. 
It is, however, noteworthy that the clear alignment between the direction of tongues (original and false) with that of the core, would have made this AR to be considered by any tilt-angle computation algorithm as a single FR, at almost any time of its emergence, if observations at only one time would have been analyzed.

%{\S\bf --- AR 9748- New emergences- Magnetograms}\\
The four tilt estimations for AR 9748 are shown in \fig{9748_tilt}g. The values of $\phia$ (black continuous curve) are positive as expected from Joy's law but its variation changes from increasing values to decreasing ones by 22 December at $\approx$ 17:30 UT, close to the time when the distribution of the tongue flux changes from positive to apparent negative twist. This change indicates first a counter-clockwise rotation and later a clockwise one. As in the other examples, CoFFE shows more stable estimation { until the end of 24 December}, which agree with Joy's law (red continuous line in \fig{9748_tilt}g). The change of position of the preceding polarity core observed during the last day (seen approximately during 2001-12-24 20:00 UT to 2001-12-25 24:00 UT in 9748\_CoFFE.mp4) is due to the increase of the flux of the second emergence, affecting also the value of $\phic$. {In fact, the correction provided by CoFFE is affected both by the spatial location and the flux of the secondary bipole relative to the first bipole.} 
% This implies that the correction provided by CoFFE depends not only in the spatial location of the secondary bipole but also on the flux strength of the core.}

%{\S\bf --- AR 9748-  New emergence- White light} \\
The blue squares and green dots in \fig{9748_tilt}g depict the evolution of $\phiwl$ and $\phiwlm$. Some dispersion exists in $\phiwl$ and $\phiwlm$, around the end of 21 December and beginning of 22 December, when the tongue umbrae are present in the first bipole (see the WL movie {9748\_WL.mp4}).
The morphology of the tongues forces a wrong grouping in the case of $\phiwl$, this is somehow corrected when including the polarity sign and $\phiwlm$ becomes closer to $\phia$. After the second bipole emergence both measurements are closer to the red curve corresponding to $\phic$, indicating that the flux in the core produces the largest umbrae.

%{S\bf --- Summary} \\
From the two examples described in this section, in the first one, AR 11007, we observe a clear new emergence in between the polarities of the first emerging bipole. This emergence produces a false rotation of the original bipole when measuring tilt values from LOS magnetograms using the magnetic barycenters, as well as wrong tilt values when using WL data. The second example, AR 9748, illustrates how a secondary flux emergence with specific characteristics can somehow trick algorithms used to compute tilt angles both from LOS magnetograms and WL data, i.e. all of them would have considered in this case the emergence of a single FR at any time. This secondary emergence also produces a false rotation of the AR main bipole. 
However, even if deceived by both ARs, CoFFE finds more stable and coherent results { during most of the AR emergence} since it computes tilt values using only the flux in the core centers (excluding most of the tongue and new emerging flux).

%--------------- Summary and conclusions -----------------------
%BEGIN CONCLUSIONS
\section{Summary and Conclusions}
%\section{Summary and conclusions}
 \label{sect_Conclusions} 

%{\S\bf --- Summary of the problem \& aims } \\
The correct determination of the tilt angle of active regions (ARs) is fundamental to understand the underlying processes that take place during the transit of magnetic flux ropes (FRs) through the convection zone.
Moreover, flux-transport dynamo models rely on the precise estimation of the latitudinal dependence of the tilt angle, known as Joy's law, to predict the passage from one solar cycle to the next \citep[see][]{Nandy18,Cameron10}. 
Tilt-angle values, and consequently Joy's law, have shown significant variation and dispersion depending both on the observable and the method used to measure them.
In this article,  we test four different methods to measure AR tilts.
We also explore the implications on the measured tilt of typical characteristics of flux emergence, namely the
evolution of magnetic tongues and the emergence of secondary bipoles.

%{\S\bf --- Summary using magnetograms: MB } \\
A standard method is to use line-of-sight (LOS) magnetograms, and to compute the tilt angle, $\phia$, 
from the flux-weighted centers (or magnetic barycenters) of the magnetic polarities (\sect{Mod_MAG}). 
However, \citet{Poisson16} have shown that the elongation of the polarities produced by the magnetic tongues can affect significantly the position of magnetic barycenters during the emergence of ARs and, consequently, the value of $\phia$.
In \sect{Strong_tongues} we have shown that the magnetic tongues can produce spurious rotations of the AR bipole and values of $\phia$ which oppose to Joy's law.
Therefore, these measurements can contribute to increase the dispersion found in statistical studies of Joy's law in which the stage of the AR evolution is not taken into account.

%{\S\bf --- Conclusions with pure WL data } \\
The earliest and largest databases used to compute tilt-angle values are those based on WL images. 
Then, these databases have been the ones mostly used in statistical studies of Joy's law \citep{Howard90,Baranyi15,Wang15}. Tilt values obtained from WL data depend, first, on the method to identify umbra areas, and second, on the algorithm used to assign each umbra to the corresponding magnetic polarity (see \sect{Mod_WL}). 
We find that the tilt values, $\phiwl$, derived using from SOHO WL images in SDD \citep{Gyori10} could differ from those derived by us using the k-means clustering method (\eg\ \figs{9574_tilt}{9906_tilt}). 
Furthermore, as recognized before \citep{Baranyi15}, the WL data include ARs which have all their spots with the same magnetic polarity. 
Then, in these unipolar regions, any clustering algorithm define wrongly a tilt angle.  
This implies a strong bias on ARs which are in their early stage of emergence and/or posses low magnetic flux. 

%{\S\bf --- Conclusion on WL with B information } \\
The above problems of $\phiwl$ can be detected using the information of the magnetic field sign to properly group the umbrae and obtain values that we call $\phiwlm$.
However, the use of WL data cannot guarantee that the effect of the magnetic tongues is removed, since in some ARs we found that the flux associated to the tongues can still produce large umbrae (\eg\ \fig{example}).
Therefore, WL tilt values have the same problems as $\phia$ in ARs with strong magnetic tongues independently of the grouping algorithm used ({\eg} \figs{9574_tilt}{9906_tilt}).  All these imply that the most common methods used to compute tilts cannot, in general, give precise estimations of tilt angles in young ARs. 

%{\S\bf --- Summary using CoFFE } \\
In \citet{Poisson20} we developed a method, called Core Field Fit Estimator (CoFFE), based on the identification of the LOS field distribution {and designed to isolate} the axial field {of the emerging} FR (\ie\ the core flux). This method allows us to eliminate the effect of magnetic tongues on tilt measurements, as well as the presence of secondary emergences, then to obtain corrected tilt values, {called $\phic$ (}\sect{CoFFE}).  
We test the consistency of $\phic$ in cases where the magnetic tongues are weak (low magnetic flux) and, {for bipolar ARs} we find no significant differences between the values achieved with the other three methods (\figs{11027_tilt}{10879_tilt} in \sect{Weak_tongues}).
In cases where the tongues are strong we find that CoFFE effectively reduces the flux associated to the magnetic tongues from the tilt estimation removing the spurious rotation of the bipole, as well as the deviation of the tilt from Joy's law predictions (Figs.~\ref{fig_9574_tilt}, \ref{fig_9906_tilt} and \ref{fig_8760_10268_tilt}).

{ We compare the performance of CoFFE with the standard methods to estimate the tilt angle for the particular cases of multipolar ARs. Although the tilt angle is only defined for ARs formed by the emergence of a single FR, most of the statistical studies use standard methods without considering the {AR characteristics}, which may lead to a larger dispersion {of tilt angle values and/or inconsistent results.
The CoFFE method can still estimate the tilt angle of the main bipole in regions with multiple emergences, {provided that} these emergences are located within the main AR bipole, {\ie} in the band defined between the main bipole polarities that is excluded by CoFFE by method design (to minimize the effects of magnetic tongues). 
Presently, we cannot generalize the application of CoFFE to all multipolar ARs, since the correction achieved depends on the spatial location and the flux strength of the secondary emergences.  Nevertheless,} we find that for the analyzed ARs CoFFE significantly {reduces} the effect of secondary flux emergences on $\phic$ (see \sect{Multipolar}).
This implies that CoFFE can be used to improve the estimation of the tilt angle in studies using large samples with statistical purposes.
{In fact, in order to treat more correctly multipolar ARs, CoFFE will need to be improved to include an algorithm which first identifies, and then separates, different emerging bipoles in a similar way as done by \citet{Leka96}.  ARs formed by a series of significant emergences will ultimately have tilt angles associated to each identified {bipole}.}  
}

%{\S\bf --- CoFFE as most reliable method } \\
Summarizing, the aforementioned standard methods, using either WL data or LOS magnetograms or a combination of both, to measure tilt angles strongly depend on the stage of the AR evolution, being the presence of magnetic tongues the main problem that affects tilt-angle estimations, {during the emerging phase}.
That is why CoFFE is designed to correct their effect. 
However, to correctly apply CoFFE, we need at least one magnetogram along the AR evolution in which the core region can be detected and isolated from the tongues in both polarities. Therefore, each AR has to be treated individually if we want to extend the computation of tilt angles as far as the early stages of the AR emergence. This more involved treatment of the data somehow reduces the applicability of the method in automatic procedures that deal with a large number of cases. Despite this limitation, we still find that CoFFE is the only method giving the most precise tilt-angle values {during} an AR evolution. 
This further allow to study how emerging ARs are rotating and to further study its physical origin (\eg\ due to the writhe of the FR axis or to the action of a convective vortex). This will extend previous studies done on the long-term evolution of ARs \citep[\eg\ ][]{Lopez-Fuentes03} to the emerging phase, with the potential to reveal more information on the sub-photospheric FRs.

%%%%%%%%%%%%%%%%%%%%%%%%%%%%%%%%%%%%%%%%%%%%%%%%%%%%%%%%%%%%%%%%%%%%%%%%%%%
%% Acknowledgements
%
 \begin{acknowledgements}

{The authors thank the anonymous reviewer for very useful comments and suggestions.} MP, MLF, and CHM acknowledge financial support from Argentine grants PICT 2016-0221 (ANPCyT) and UBACyT 20020170100611BA. MLF and CHM are members of the Carrera del Investigador Cient\'{\i}fico of the Consejo Nacional de Investigaciones Cient\'{\i}ficas y T\'ecnicas (CONICET). MP is a fellow of CONICET. {This work was supported by the Programme National PNST of CNRS/INSU co-funded by CNES and CEA.} The authors acknowledge the use of data from the SOHO (ESA/NASA) mission. These data are produced by the MDI international consortia.

 \end{acknowledgements}

\bibliographystyle{aasjournal}  % format of ref. provided by the review (.bst)
\bibliography{paper_tongues}  % file containing the bibtex references (.bib)
     %  look if the file containing the ``\bibitem'' exits
\IfFileExists{\jobname.bbl}{}
{\typeout{}
\typeout{****************************************************}
\typeout{****************************************************}
\typeout{** Please run "bibtex \jobname" to obtain}
\typeout{** the bibliography and then re-run LaTeX}
\typeout{** twice to fix the references!}
\typeout{****************************************************}
\typeout{****************************************************}
\typeout{}
}

\begin{appendix}

%--------------------------------------------------------------
\section{A Variety of Tongue Morphologies and Evolutions}
%\section{A variety of tongue morphologies and evolutions}
\label{app_Variety_tongues}

%{\S\bf --- Generalities on tongue variety} \\
\citet{Poisson16} studied the characteristics of magnetic tongues for 149 bipolar ARs  observed along a full solar cycle. Though, in general, tongues have a tendency to be stronger at the start of the emergence and become weaker as the magnetic flux of the AR reaches its maximum, there are many cases in which tongues stay strong and extended even at the time of 
maximum flux (see the examples in \sect{Strong_tongues}). Furthermore, observed tongues present a large variety of morphologies and evolutions, even appearing at any stage of an AR emergence. This large variety could be only reproduced  when  using  a  broad  range  of  twist profiles when comparing {the data with the emergence of a twisted FR.  This is shown in Figure 10 of \citet{Poisson16} where} observations are compared with the results of FR models with varying radial and azimuthal twist profiles. In this section we discuss two examples in which tongues evolve differently; they appear, develop, and almost disappear in AR 10268, while they are mostly present and very elongated during the full evolution of AR 8760.

\begin{figure}[!ht]
\begin{center}
\includegraphics[width=.45\textwidth]{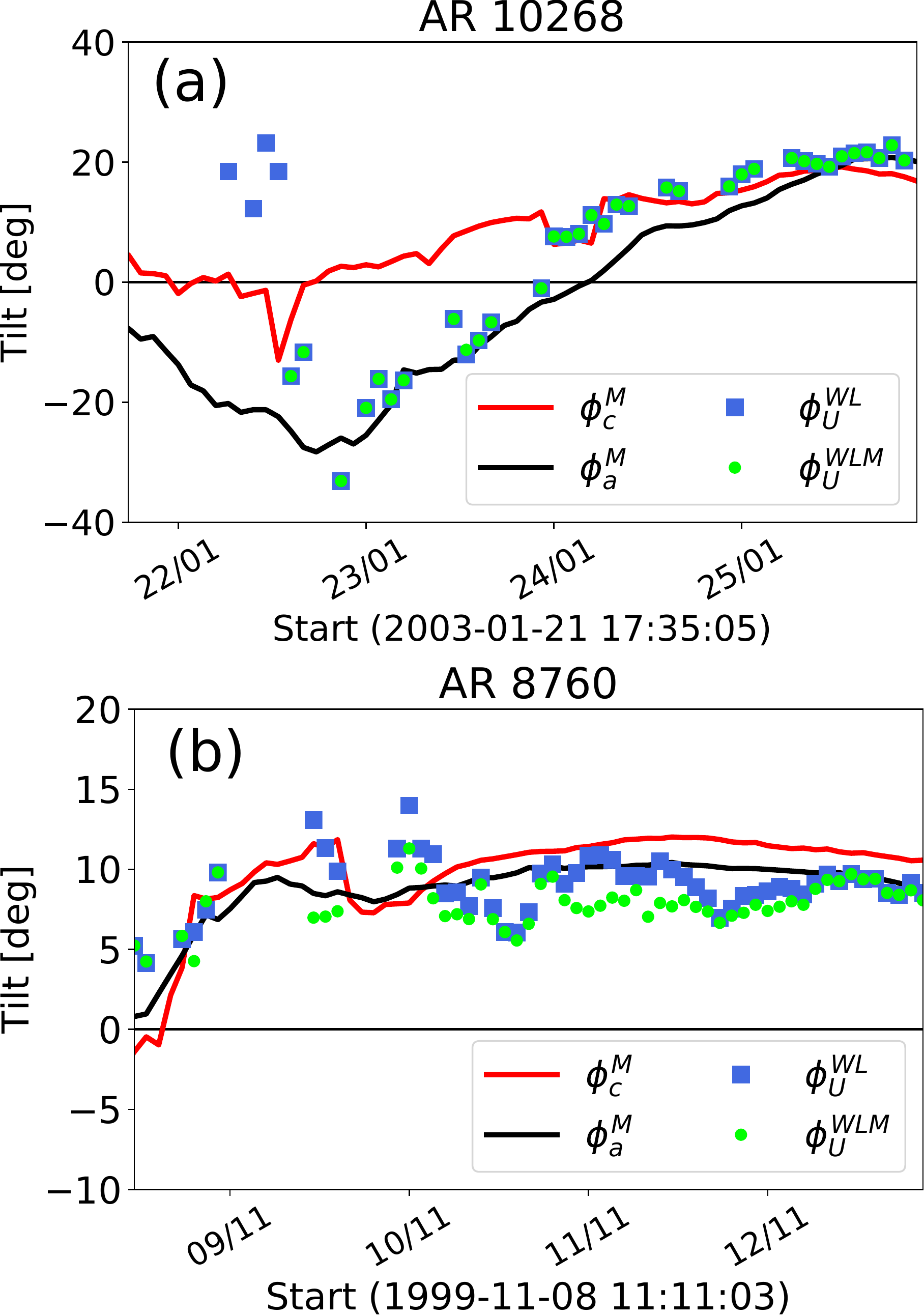}
\caption{Evolution of the tilt angles along the emergence of (a) AR 10268 and (b) AR 8760, both located in the northern hemisphere.
The black and red continuous lines, as well as the blue squares and green dots, have the same meaning as those in \fig{11027_tilt}c.
{Movies of these AR evolutions are available as additional material (10268\_WL.mp4, 10268\_CoFFE.mp4, 8760\_WL.mp4 and 8760\_CoFFE.mp4).}
}
\label{fig_8760_10268_tilt}
\end{center} 
\end{figure}

%{\S\bf --- AR 10268 - evolving strong tongues - Magnetograms } \\
 AR 10268, which emerges in the northern hemisphere, has a clear bipolar configuration in which tongues are strong and do not appear clearly separated from the core flux (see the LOS magnetic field movie 10268\_CoFFE.mp4). The effect of tongues is evident when comparing the black and red curves in \fig{8760_10268_tilt}a. The black curve, $\phia$, shows that AR 10268 evolves as it emerges towards a high tilt value that opposes to Joy’s law ($\phia<0$), while by the beginning of 23 January 2003 there is a sudden change in the apparent bipole rotation towards tilt values agreeing with this law ($\phia>0$). This variation implies that the bipole would first rotate counter-clockwise by around 20$^\circ$ and later clockwise by around 45$^\circ$. A different tilt evolution is shown by the red curve depicting the values of $\phic$ {computed with CoFFE}. In this case, most values agree with what is expected from Joy's law and the evolution of the corrected tilt angle indicates a consistent clockwise rotation of $\approx 20^\circ$. {Summarizing, in} this AR we observe a typical behavior of the tongues, i.e. they appear in the first stages of the emergence, evolve, and almost disappear by its end; this is shown by the coincidence between $\phia$ and $\phic$ in \fig{8760_10268_tilt}a at the end of the emergence phase. 

%	FIGURE A.1  ---------------------------------------------
%--------------------------------------------------------------  

%{\S\bf --- AR 10268 - evolving strong tongues - White light}\\
Concerning tilts derived from WL data, \fig{8760_10268_tilt}a {shows} that the blue squares, $\phiwl$, and the green dots, $\phiwlm$, are quite scattered and do not clearly follow either the black or red curves. 
In this AR, tongues are so strong that they have umbrae and they affect the WL tilt measurements. Next, we observe four wrong $\phiwl$ values derived from unipolar umbra measurements (as in the case of AR 10879, \fig{10879_tilt}). After the beginning of 23 January both, $\phiwl$ and $\phiwlm$, follow roughly the evolution of $\phia$ until the time when tongues start retracting on 24 January. By this time, both WL measurements follow the evolution of $\phic$.  Finally, at the end of the emergence all four tilt values agree (see the WL movie 10268\_WL.mp4). 

%{\S\bf --- AR 8760- varying intensity tongues- Magnetograms } \\
AR 8760 emerges in the northern solar hemisphere. This is a mainly bipolar AR which shows a series of minor emergences in between the two main bipole polarities almost all along the period of time shown in \fig{8760_10268_tilt}b (see the movie 9760\_CoFFE.mp4).  
{ These minor bipoles can be intepreted as the resistive emergence of an undulatory FR in which the upper part is fragmented by a Parker instabiliy and then reformed by magnetic reconnection at the photospheric layer \citep{Pariat04,Cheung10}. Despite its complex evolution, these minor bipoles do not affect much tilt-angle measurements}. At the beginning of the emergence tongues are not clearly visible due to the presence of a secondary bipole.  Next, by 8 November 1999 at $\approx$ 19:10 UT a clear elongated tongue pattern, corresponding to a negatively twisted FR, is present.  
After a fast increase, a comparable evolution of $\phia$ and $\phic$ (clockwise rotation) is present. 
The largest difference, of around 5$^\circ$ in $\phic$ above $\phia$, is evident at $\approx$ 14:25 UT on 9 November.
By the beginning of 11 November, tongues are smaller and weaker, though still present.  
It is this distribution of the flux what now makes both $\phia$ and $\phic$ follow the same behavior until the end of the period shown in \fig{8760_10268_tilt}b. The values of the tilt angles derived from magnetograms agree with what is expected from Joy's law. 

%{\S\bf --- AR 8760- varying intensity tongues- White light } \\ 
Umbrae are present in the core regions as well as in the tongues once they become clearly visible (see the WL movie, {8760\_WL.mp4}, notice also that there are several gaps in these data). The blue squares and green dots, $\phiwl$ and $\phiwlm$, follow the increase in tilt-angle values as $\phia$ and $\phic$ in \fig{8760_10268_tilt}b. At around the time of the largest difference between $\phia$ and $\phic$, $\approx$ 14:25 UT on 9 November, umbrae are very disperse and the k-means algorithm fails to group them correctly locating some of them on the wrong polarity region giving the largest difference between the blue squares and green dots. This failure has in fact a positive effect since it decreases the effect of the tongues, as in AR 9906 (\fig{9906_tilt}), then  $\phiwl$ values are closer to the red continuous curve of $\phic$ at that time.  After a gap in the WL data, ending at around the beginning of 10 November, both WL measurements follow the same behavior with differences between them of less than 5$^\circ$. The largest differences happen when k-means clustering groups umbrae located in the wrong polarity sign region. Later on both WL tilt values, $\phiwl$ and $\phiwlm$, stay closer to the black continuous line corresponding to $\phia$, showing again the effect of magnetic tongues.     

%{\S\bf --- Summary}\\
This section shows how diverse, both in morphology and evolution, tongues can be. 
In AR 10268 tongues are strong enough at the start of the emergence to affect
the apparent tilt evolution as well as measurements using WL data. It is only when they start to retract that a fair agreement of the four tilt-angle values is observed. 
In the second example, AR 8760, tongues are present all along the emergence with a varying intensity.
WL measurements are clearly affected by the dispersion of the umbrae that are present both in the core and tongue regions. Despite some differences between $\phiwl$ and $\phiwlm$, mainly caused by a wrong grouping, they follow in general the tilt values given by the magnetic barycenters method.       

\end{appendix}

\end{document}